\documentclass{article}
\usepackage{ijcai25}

\usepackage{times}
\usepackage{soul}
\usepackage{url}
\usepackage[hidelinks]{hyperref}
\usepackage[utf8]{inputenc}
\usepackage[small]{caption}
\usepackage{graphicx}
\usepackage{amsmath}
\usepackage{amsthm}
\usepackage{booktabs}
\usepackage{algorithm}
\usepackage{algorithmic}
\usepackage[switch]{lineno}

\newcommand{\lightgray}[1]{{\color{lightgray} {#1}}}

\usepackage{graphicx} % Required for inserting images
\usepackage{natbib}
\usepackage{verbatim}
\usepackage{amsmath,amssymb,amsfonts,bm}
\usepackage{bbm}
\usepackage{enumitem}
\usepackage{amsthm}
\usepackage{textcomp}
\usepackage{float}
\usepackage[font=small]{caption}
\usepackage{xcolor}
\usepackage{wrapfig}
\usepackage{hyperref}
\usepackage[font=footnotesize,labelformat=simple]{subcaption}
\usepackage[protrusion=true,expansion=true]{microtype}
\pdfoutput=1

\usepackage{caption}
\usepackage{multirow}
\usepackage{graphicx}
\usepackage{slashbox}
\usepackage{amsthm}

\usepackage[font=footnotesize,labelformat=simple]{subcaption}
\everymath{\displaystyle}

\newcommand{\satya}[1]{{\color{red} {Satya: #1}}}

\title{Physics-based Generative Models for Geometrically Consistent and Interpretable Wireless Channel Synthesis}

\author{
Satyavrat Wagle$^1$\and Akshay Malhotra$^2$\and Shahab Hamidi-Rad$^2$ \and Aditya Sant$^2$\and David J. Love$^{^1}$ \And Christopher G. Brinton$^{^1}$
\affiliations
$^1$Elmore Family School of Electrical and Computer Engineering, Purdue University\\
$^2$InterDigital Communications\\
\emails
wagles@purdue.edu,
firstname.lastname@interdigital.com,
\{djlove,cgb\}@purdue.edu
}

\begin{document}

\maketitle

\begin{abstract}
    In recent years, machine learning (ML) methods have become increasingly popular in wireless communication systems for several applications. 
    A critical bottleneck for designing ML systems for wireless communications is the availability of realistic wireless channel datasets, which are extremely resource-intensive to produce. To this end, the generation of realistic wireless channels plays a key role in the subsequent design of effective ML algorithms for wireless communication systems. 
    Generative models have been proposed to synthesize channel matrices, but outputs produced by such methods may not correspond to geometrically viable channels and do not provide any insight into the scenario being generated. 
    In this work, we aim to address both these issues by integrating established parametric, physics-based geometric channel (PPGC) modeling frameworks with generative methods to produce realistic channel matrices with interpretable representations in the parameter domain.
    We show that generative models converge to prohibitively suboptimal stationary points when learning the underlying prior directly over the parameters due to the non-convex PPGC model. To address this limitation, we propose a linearized reformulation of the problem to ensure smooth gradient flow during generative model training, while also providing insights into the underlying physical environment.
    We evaluate our model against prior baselines by comparing the generated, scenario-specific samples in terms of the 2-Wasserstein distance and through its utility when used for downstream compression tasks.
\end{abstract}

\section{Introduction}\label{sec:intro}
\iffalse
\satya{\begin{itemize}
    \item ML applications for the wireless domain such as channel estimation, compression, prediction, beamforming, etc. (Refs for all) require a large amount of wireless data
    \item Collecting, filtering and cleaning this data is complex, tedious and time consuming as it requires human intervention for labelling.
    \item Generative models have been proposed as a method to synthesize channel data for downstream applications (Ref.)
    \item Intro of generative models - focus on VAEs.
    \item However, these models suffer from two issues. (i) Synthesized outputs may not correspond to valid channel data and (ii) the outputs do not provide any insight into the scenario of interest.
    \item To address this issue, we propose to integrate established PPGC modelling frameworks into the generative process.
    \item This ensures that generated channels are valid and as the framework is parametric, insights related to the scenario can be extracted from the parameters associated with the generated channels.
    \item List of contributions.
\end{itemize}}
\fi

The use of machine learning (ML)  for applications in wireless communication has seen extensive interest in the past few years. At the physical layer (PHY) of wireless communication systems, ML research has predominantly focused on two main objectives: estimating and mitigating distortions in electromagnetic signals during over-the-air (OTA) transmission (such as channel estimation, channel compression, equalization, and beamforming),\cite{ch_compression,ch_estimation,beamforming,dl_wireless_survey,sant2022deep} and addressing noise and non-linearities at transmitting or receiving antennas \cite{estimation_survey,sant2024insights,nguyen2021dnn}. For practical PHY layer deployments, the ML pipeline requires a substantial amount of OTA wireless channel data to effectively train the models. However, the process of manually collecting, cleaning and labeling wireless data from the real world is often complex and expensive both in terms of resources and time. Past data measurement and labeling campaigns have taken multiple months to capture a handful of fully characterized data points for a single scenario \cite{OTAdata_capture1,OTAdata_capture2,OTAdata_capture3}. 

Capitalizing on the recent advances in AI, generative models have been proposed to mitigate this problem by artificially synthesizing wireless data \cite{channelgan}, significantly reducing the effort required to create wireless channel datasets for the aforementioned applications. However, unlike common modalities of data that we typically encounter, such as image, text, audio, etc. which are directly human-interpretable, the wireless channel data is a tensor of complex numbers and is \emph{not human-interpretable or easily visualized}. Combined with the inherently stochastic nature of generative models, this poses two major challenges around effectively testing and using generative channel models.  
%generative models are inherently stochastic in nature, and hence suffer from two major problems. 
Firstly, the outputs of generative models may not correspond to valid channels.  Here, the validity of channel data implies that the wireless channel can be represented as a multipath geometric model representing the multiple paths the transmitted signal takes, before arriving at the receiver \cite{chanmodel_1,chanmodel_2}. While the generative models are trained to generate data points statistically similar to the training set, owing to the stochasticity of the model, the synthesized outputs may not correspond to a geometrically valid set of interactions or multipaths.
Secondly, it is hard to gain any insights about the physical parameters associated with the signal propagation or any information about the environment or scenario being considered (e.g. angles associated with paths, gains of paths, line-of-sight transmission or non-line of sight, etc.) from generated data samples.
%\textcolor{blue}{Thirdly, construction of channel matrices from channel parameters depends on the tractability of such models. Statistical methods for channel generation, relying on tractable channel parameter distributions, cannot be specialized to specific channel scenarios where multi-parameter correlations cannot be easily characterized. However, the use of generative channel models enables learning such correlation, going beyond statistical channel generation, to create channels tailored to specific physical environments, operating conditions, user nonlinearities, etc.}

This work primarily focuses on the design of generative models for synthesizing millimeter wave (mmWave) channels, which forms the backbone of next-generation wireless communication and IoT systems \cite{iot_mmwave1,iot_mmwave2}.
The proposed method to generate mmWave channels overcomes the limitations of existing approaches by incorporating a verified PPGC model into the generative pipeline. As the PPGC model parametrizes the channel generation, our generative process learns the joint distribution of the underlying parameters responsible for channel generation. %\textcolor{blue}{Further, the proposed approach is able to generate correlated channel parameters, without the need to explicitly model the inter-parameter correlations. This is particularly significant for real-world applications to tailor the channels to the environment and application of interest.}

This mitigates both the above-mentioned issues, as by incorporating a verified PPGC model in the pipeline, the outputs of our framework are guaranteed to be valid channels, and as our model generates the parameters associated with the channels, we can extract insights related to the environment in which the channels were recorded, making the channels generated by our system more interpretable.

We further discuss the challenges associated with training the generative model with the PPGC model in the loop due to the non-convex loss landscape induced by the PPGC model. We further propose a linearized reformulation of the PPGC model, which mitigates these issues, and enables the integration of the PPGC model in the generative pipeline.

%Further, we discuss the severe non-convexity of the physics-based PPGC model with respect to its parameters, which prevents the flow of gradients through the PPGC model, resulting in updates to the generative model being negligible. In order to address this, we linearize the physics based PPGC model via the construction of an array response dictionary and assigning weights to each element, as we will discuss in Sec. \ref{sec:pred_matrix}. Additionally, we encourage sparsity in the weights given to the array response dictionary, resulting in only those elements chosen which correspond to propagation paths between the transmitter and receiver. We utilize a lightweight parameter estimation module on top of the generative process to extract parameter values which correspond to these paths. 

\subsection{Summary of Contributions}
The contributions of our paper are as follows.
\begin{itemize}
    \item We propose a generative ML framework which leverages a parametric, physics-based channel (PPGC) model to produce realistic channel data that belongs to the distribution of interest (Sec. \ref{sec:model}).
    \item We explore the challenges associated with the training of the proposed generative framework arising from the non-convexity of the PPGC model (Sec. \ref{sec:pred_params}). 
    \item We develop a linearized relaxation of the PPGC model to mitigate the effect of this non-convexity (Sec. \ref{sec:pred_matrix}).
    %\item We develop a parameter estimation module that calculates the parameters associated with a generated sample of data, allowing for a higher degree of interpretability of the outputs as well as the extraction of environment-related insights (Sec. \ref{sec:parameter_estimation}).
    \item We show that our method can accurately generate channel data as well as the parameter distributions associated with a given set of real data (Sec. \ref{sec:results}).
    \item We evaluate our method against prior arts baselines, and experimentally show that our method is able to capture scenario-specific distributions more accurately (Sec. \ref{sec:results}).
\end{itemize}

\section{Related Work}\label{sec:rel_work}
\iffalse
\satya{\begin{itemize}
    \item ChannelGAN
    \item Score based GANs
    \item MIMOGAN
    \item Diffusion models for PPGC modelling
\end{itemize}}
\fi

Several works propose using a generative model to produce novel channel samples through the stochastic generative process. A generative  adversarial network (GAN) based wireless channel modeling framework was first introduced in \cite{GAN1}. \cite{channelgan} utilize a Wasserstein-GAN with Gradient Penalty (WGAN-GP) to synthesize novel channel matrices given a limited set of training data points and are evaluated by cross-validating between real and synthesized data. \cite{mimogan} trained their model on multiple-input multiple-output (MIMO) data, with a discriminator explicitly designed to learn the spatial correlation across the channel data.
In \cite{diffusion_models_for_channels}, a diffusion based generative model has been adopted to circumvent the issue of mode collapse in GANs. They further evaluate the overlap of generated data with real data by calculating the approximate 2-Wasserstein distance between the power spectra. Works such as \cite{score_based} utilize a score-based generative model to generate channel matrices and further extend the framework for channel estimation in noisy environments. All of the above works utilize a generative model to directly produce channel matrices at the output, but they have no guarantees on the validity of the generated channel data, with limited interpretability. %In contrast, our method utilizes the generative model to produce a distribution over the parameter manifold, which is utilized by a verified channel model to produce the channel matrix.

A similar research direction involves using labeled datasets to predict parameters associated with the wireless channel. In \cite{gen_models_mmwave_uav}, the authors use the locations of UAVs to predict the link state and the channel parameters sequentially, using a conditional variational autoencoder (VAE) to capture complex statistical relationships within the data. Similar to the previous work, in \cite{multi_freq_model} the authors develop a GAN based model to generate new instances of channel parameters given the location of UAVs. 
%In \cite{spatially_consistent_a2g}, the authors use labeled trajectories of RSS values to predict similar, spatially consistent trajectories of RSS. 
The above works require datasets labeled with metadata relating to the environment, locations of the transmitter or receiver, and the entire set of channel parameters, and can not evaluate the channel matrices directly. In contrast, our method does not require labeled data and can learn channel parameters directly from the channel matrix.

\section{System Model and Approach}\label{sec:model}

In this section, we describe our proposed system. We begin by describing the PPGC model that is used in the generative process, followed by the design of the generative model. We discuss the design choices made for the generative model. Finally, we describe the training and inference pipeline.

\subsection{Channel Model}\label{sec:channel_model}
We consider a wireless communication system with $N_t$ transmit and $N_r$ receive antennas. The associated PPGC model defined by $\mathcal{M}: \mathbb{R}^{3P} \rightarrow \mathbb{R}^{2 \times N_t \times N_r}$ \cite{pb_model}, maps a set of parameters $\textbf{s} \in \mathbb{R}^{3P}$ to a matrix $\textbf{H} \in \mathbb{C}^{N_t \times N_r}$, where $P$ is the number of paths that a transmitted signal takes before being received at the receiver antennas. In mmWave channels, the total number of paths $P$ is typically small in over-the-air transmission. The PPGC model considers the channel matrix $\textbf{H}$ to be a superposition of the individual propagation matrices associated with each of the $P$ paths.

\begin{equation}
\label{eq:channel_sum}
    \textbf{H} = \mathcal{M}(\textbf{s}) = \sum_{p=1}^{P} g_p \textbf{a}_r(\theta^p_a) \textbf{a}_t(\theta^p_d)^H.
\end{equation}

Where, $\textbf{s} = [g_p,\theta_a^p,\theta_d^p]_{p = 1}^P$,
%set of parameters $\textbf{s}$ for the $p$-th path indexed consists of 
$g_p$ represents the propagation gain associated with the $p$-th path, $\textbf{a}_t(\cdot), \textbf{a}_r(\cdot)$ represent the steering vectors or the array response vectors on the transmit and receive antennas, and $\theta_d^p$ and $\theta_a^p$ represent the corresponding angle of departure and angle of arrival, both of which take values between $[-\pi,\pi]$ radians. To simplify the discussion, we consider a Uniform Linear Array (ULA) \cite{chanmodel_1} and limit the discussion to the azimuth plane. Thus, the array response vectors can be defined as:
\begin{align}
\label{eq:array_responses}
    \textbf{a}_t(\theta_d^p) = \frac{1}{\sqrt{N_t}}[1,e^{ju\sin(\theta_d^p)},\dots,e^{j(N_t-1)u\sin(\theta_d^p)}]^T,\\
    \textbf{a}_r(\theta_a^p) = \frac{1}{\sqrt{N_r}}[1,e^{ju\sin(\theta_a^p)},\dots,e^{j(N_r-1)u\sin(\theta_a^p)}]^T,
\end{align}

%$\theta^a_p$, angle of departure $\theta^d_p$ and the path gain $g_p$. We assume that the number of paths $P$ is known.

%Internally, the channel model $\mathcal{M}$ first calculates the antenna array responses for the transmit and receive antennas based on the angles of departure and arrival, given by $\textbf{a}_r(\theta^a_p)$ and $\textbf{a}_t(\theta^d_p)$ respectively. Assuming Uniform Linear Arrays (ULA), the array response vectors are calculated as follows

where, $u = \frac{2\pi}{\lambda}d$, $\lambda$ is the wavelength of the carrier and $d$ is the distance between antenna elements. 
%We consider the DFT of the channel matrix as it is a sparse representation of the channel \cite{group_sparsity}, which restricts the generated channels to sparse matrices, thus aiding in downstream channel generation.The output DFT channel matrix is calculated as follows

%\satya{Check theta subscripts.}

%where $\textbf{D}_R^H$ and $\textbf{D}_T$ are DFT matrices of size $N_r \times N_r$ and $N_t \times N_t$ respectively.
A key characteristic of the PPGC model $\mathcal{M}$ is that, a distribution $q(\cdot)$ over the parameters $\textbf{s}$ induces a distribution $q_M(\cdot)$ over the channel matrix $\textbf{H}$, where the distribution $q(\cdot)$ is specific to the environment in which the model is deployed. Now, in order to incorporate the PPGC model into the generative pipeline, the generative model must learn the prior distribution over the parameters $\textbf{s}$ instead of directly learning the prior distribution over the channel matrix $\textbf{H}$. Thus, the objective of our system is as follows; given a set of channel matrices $\mathcal{D} = [\textbf{H}_i]_{i=1}^N$, where $\textbf{H}_i \sim q'(\cdot)$ and $q'(\cdot)$ is unknown, we train a generative model to output a probability distribution $q(\cdot)$ over the parameters $\textbf{s}$ such that the induced distribution $q_M(\cdot)$ over the channel $\textbf{H}$ is close to $q'(\cdot)$.
%log-likelihood function {\tiny{$\sum_{i=1}^N$}} $q_{M}(\textbf{H}_i)$ is maximized, which is equivalent to minimizing the KL divergence between $q_M$ and $q'$ \cite{mle_kld}.} 
%\satya{I'm not sure if this is the right way to say this. If it's tedious, we can just remove this and let the earlier sentences explain.}%\satya{Change here. Show multiple solutions for Jacobian? Training loss graph?}

\begin{figure}
    \centering
    \includegraphics[width=0.98\columnwidth]{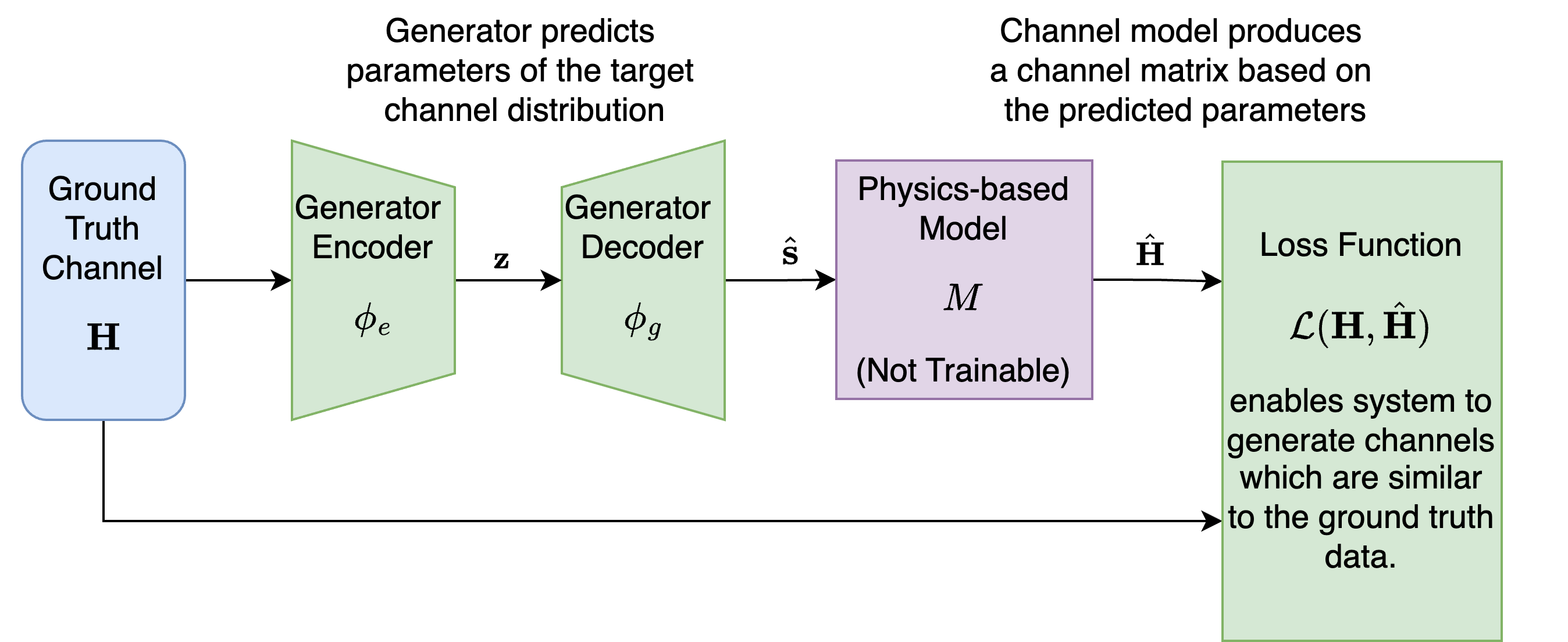}
    \includegraphics[width=0.8\columnwidth]{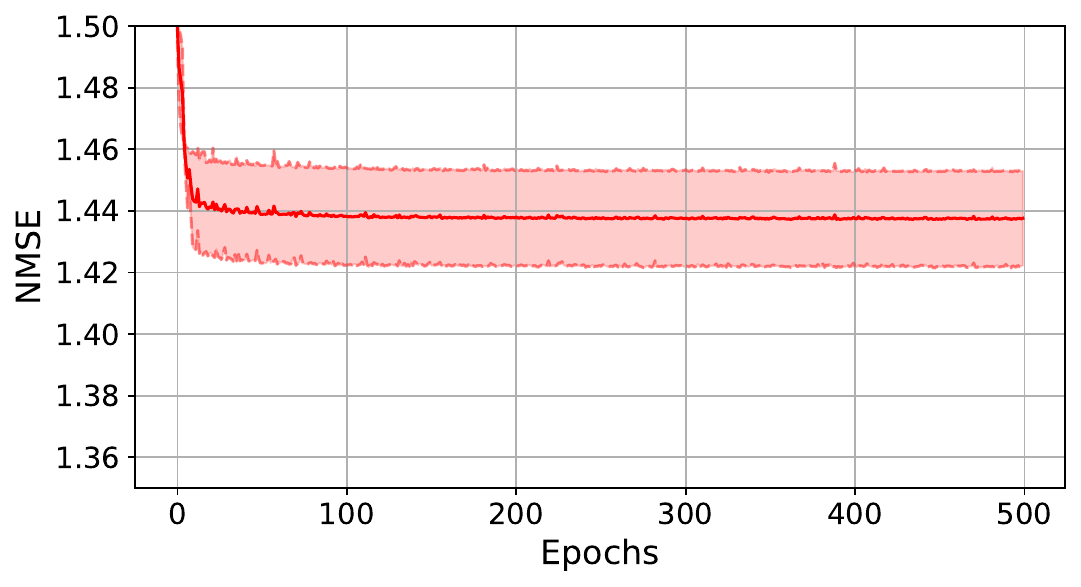}
    \caption{When integrating the PPGC model in a straightforward manner, the generator directly predicts the parameters $\hat{\textbf{s}}$, which are then used by the PPGC  model $\mathcal{M}$ to produce the predicted channel (Top). In such implementations, the generator cannot converge to suitable optima due to the non-convexity of the PPGC model, as observed in the training performance of the pipeline over multiple instances. (Bottom)}.
    \label{fig:model_train}
\end{figure}

\subsection{Generative Model to Predict Channel Statistics} \label{sec:pred_params}

For the generative model, we use the variational autoencoder (VAE) architecture \cite{vae}, as seen in Fig. \ref{fig:model_train}. We use the generative model to produce the parameter vector $\textbf{s}$ using a latent variable $\textbf{z}$, which is then passed to the PPGC model $\mathcal{M}$ to produce a valid channel matrix. 
%The generative model consists of an encoder $f_{\phi_e}:\mathbb{R}^{2 \times N_t \times N_r}\rightarrow \mathbb{R}^{Z}$ and a decoder $g_{\phi_d}:\mathbb{R}^{Z} \rightarrow \mathbb{R}^{3P}$ which are parametrized by $\phi_e, \phi_d$ respectively.
 The VAE consists of a decoder network $g_{\phi_d}(\cdot)$, parametrized by $\phi_d$, which is used to reconstruct the parameter vector $\textbf{s}$ given the latent variable $\textbf{z}$ as $g_{\phi_d}(\textbf{z})$, and the encoder $f_{\phi_e}(\cdot)$, parametrized by $\phi_e$, which approximates the posterior distribution of  $\textbf{z}$ given the channel matrix $\textbf{H}$ using the variational posterior  $f_{\phi_e}(\textbf{H})$.  

\subsubsection{Generative model training}
During the training phase, the encoder of the generative model takes a channel matrix \textbf{H} as input and samples a latent vector $\textbf{z}$ from the posterior distribution as
\begin{equation}
\label{eq:vae_encode}
    \textbf{z} \sim f_{\phi_e}(\textbf{H}).
\end{equation}
The decoder of the generative model then takes in the latent vector $\textbf{z}$ as input and produces a parameter vector $\hat{\textbf{s}}$ as
\begin{equation}
\label{eq:vae_decode}
    \hat{\textbf{s}} = g_{\phi_d}(\textbf{z}).
\end{equation}
The predicted parameter vector is then passed to the model $\mathcal{M}$ to produce an output channel $\hat{\textbf{H}}$ as follows 
\begin{equation}
\label{eq:direct_param_pred}
    \hat{\textbf{H}} = \mathcal{M}(\hat{\textbf{s}}).
\end{equation}
The system loss is a generalization of the evidence based lower bound (ELBO)\cite{vae}, given by
\begin{equation}
\label{eq:vae_loss_direct}
    \mathcal{L} = ||\textbf{H}-\hat{\textbf{H}}||_2^2 + \alpha_{D} \cdot \textsf{KL}(\textbf{z},\mathcal{N}(0,\textbf{I})).
\end{equation}
Here, the first term corresponds to the reconstruction or mean square error (MSE) loss between the input $\textbf{H}$ and the predicted channel matrix $\hat{\textbf{H}}$. This ensures that the outputs are similar to the inputs. The second term penalizes the Kullback-Leibler (KL) divergence \cite{kldiv} between the latent vector $\textbf{z}$ and a simple, known distribution, in this case, the multivariate unit Gaussian distribution $\mathcal{N}(0,\textbf{I})$, where $\textbf{I}$ is the identity matrix of dimension $Z$. This encourages the distribution of the latent vectors to be similar to $\mathcal{N}(0,\textbf{I})$.

%\subsubsection{Generative model inference - PPGC generation}
Channel generation by incorporating the PPGC model proceeds in accordance with the standard VAE architecture. The loss function \eqref{eq:vae_loss_direct} enforces a latent space distribution similar to $\mathcal{N}(0,1)$, with the decoder $g_{\theta_{d}}(\cdot)$ trained as the channel generator network. The input to the decoder is a random variable $\mathbf{z}\sim\mathcal{N}(0,\mathbf{I})$. The decoder generates a set of parameters $\mathbf{\hat{s}}$, corresponding to the physics-based parameters of a new channel. These parameters are passed through the PPGC model $\mathcal{M}$ to generate an interpretable and valid channel matrix. However, the direct use of the physics-based channel model, using \eqref{eq:channel_sum} requires knowledge of the number of multipath components $P$. Further, gradient backpropagation using the PPGC model results in poor training performance, owing to the non-convexity of the model. This is further elucidated below.

\begin{figure*}
    \centering
    %\vspace{-6mm}
    \includegraphics[width=0.3\textwidth]{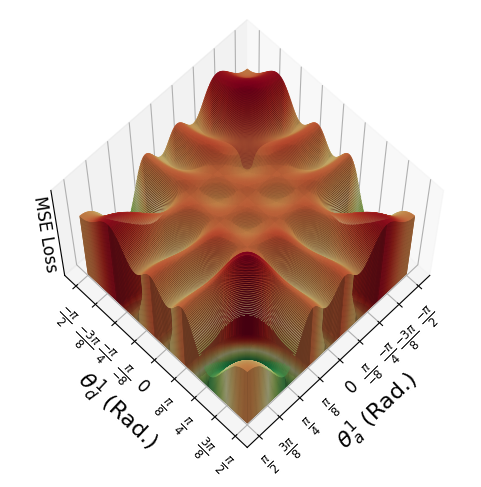}
    \includegraphics[width=0.3\textwidth]{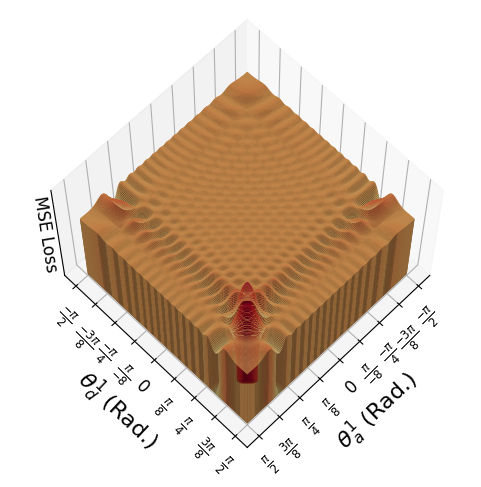}
    \includegraphics[width=0.3\textwidth]{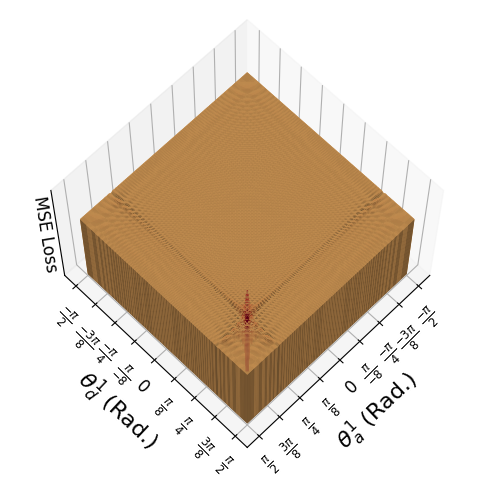}
    
    \includegraphics[width=0.3\textwidth]{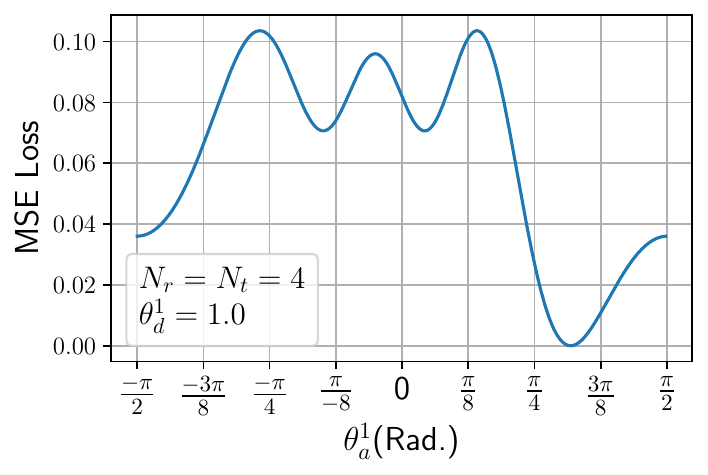}
    \includegraphics[width=0.3\textwidth]{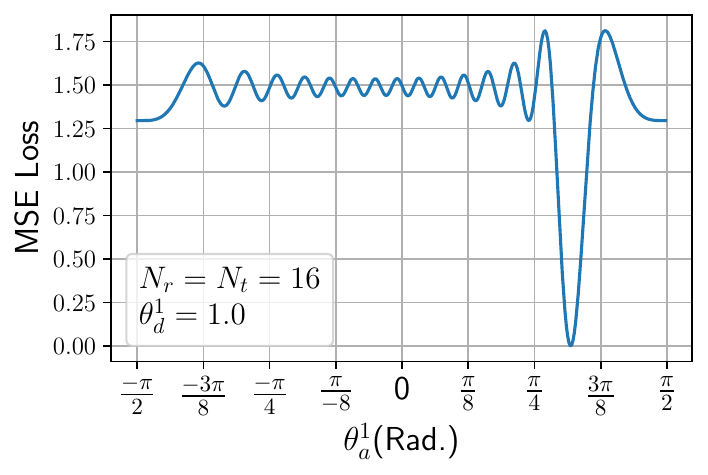}
    \includegraphics[width=0.3\textwidth]{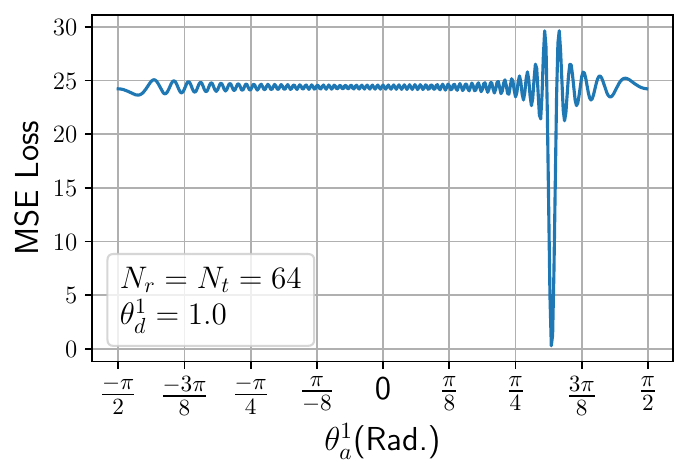}
    \caption{The loss surface as a function of $(\theta_a^1,\theta_d^1)$ in reference to a channel matrix with $\theta_a^1 = \theta_d^1 = 1.0$ radians using a PPGC model $\mathcal{M}$ with $P=1$ and $N_r = N_t = 4,16,64$ antennas respectively. The PPGC model $\mathcal{M}$ is extremely non-convex as a function of the parameters $\theta_a,\theta_d$ because of periodicity arising from the formulation of the array response vectors. Avoiding the numerous local minima surrounding the global minima poses a significant challenge given the nature of activation functions. As the number of antennas $N_r,N_t$ increases, the gradient flowing through the model diminishes significantly at locations further away from the minima, where the loss surface is effectively flat. }
    \label{fig:nonconvex_surface}
\end{figure*}

%\textcolor{blue}{Consider the partial derivatives ${\partial{\mathcal{L}}}/{\partial{\theta_a^p}}$ and ${\partial{\mathcal{L}}}/{\partial{\theta_d^p}}$. We have} 
%\begin{align*}\label{eq:partial_der}
%    \frac{\partial{\mathcal{L}}}{\partial{\theta_a^p}} = \frac{2(\textbf{H}-\hat{\textbf{H}})^H}{\sqrt{N_t}}[jku\cos(\theta_a^p)\cdot e^{jku\sin(\theta_a^p)}]_{k=0}^{N_t-1},\\
%    \frac{\partial{\mathcal{L}}}{\partial{\theta_d^p}} = \frac{2(\textbf{H}-\hat{\textbf{H}})^H}{\sqrt{N_r}}[jku\cos(\theta_d^p)\cdot e^{jku\sin(\theta_d^p)}]_{k=0}^{N_r-1}.
%\end{align*}

%\textcolor{blue}{In the above equations, we observe that the solutions for ${\partial{\mathcal{L}}}/{\partial{\theta_a^p}}=0$ and ${\partial{\mathcal{L}}}/{\partial{\theta_d^p}}=0$ are only at $\cos(\theta_a^p)=0$ and $\cos(\theta_d^p)=0$ respectively. Thus, there are infinitely many solutions for ${\partial{\mathcal{L}}}/{\partial{\theta_a^p}}=0$ and ${\partial{\mathcal{L}}}/{\partial{\theta_d^p}}=0$, which create local optima in the loss function,resulting in the convergence of the generator model to a suboptimal solution.} 

\subsubsection{Limitations of generative model training using PPGC}
In order to understand the limitations of training a generative network directly in conjunction with the PPGC model \eqref{eq:channel_sum}, we analyze the role of the ULA manifold, given in \eqref{eq:array_responses}. %using the angle parameters $\theta_{d}^{p}$ and $\theta_{a}^{p}$. 
The presence of sinusoidal functions of the angular channel parameters, $\theta_{d}^{p}$ and $\theta_{a}^{p}$, in creating the array response vector results in periodicities in the loss function landscape. The non-convexity arising from this periodicity is only exacerbated with the addition of multiple paths. 
However, these periodicities cannot be effectively approximated by the non-linear activation functions used in deep neural networks, leading to difficulties in training \cite{rectified_boltzman,guide_to_conv_aritmetic}. As a result, depending on the location of the optimizer in the parameter space, the gradient may not flow during backpropagation, resulting in the optimizer converging at non-optimal points. 
Additionally, 
%\emph{the convergence of the system to a suitable minimum is highly dependent on the initialization of the generative model parameters} and 
the convergence of the system to a suitable minimum is highly dependent on the number of antennas used in the PPGC model. This is illustrated in Fig. \ref{fig:nonconvex_surface} where, as the number of antennas, $N_t, N_r$ increase, the number of local minima also increase. Further, the optimality gap, the difference between the loss values at the local minima and the global minima, widens as the number of antennas increases, causing convergence to any local minima to significantly impact the overall loss, as observed in Fig. \ref{fig:model_train}. Also, as observed from Fig. \ref{fig:nonconvex_surface}, the loss landscape flattens as parameter values move further from the global optimum, leading to smaller gradients. This diminishes the effectiveness of optimization processes and gradients in achieving the optimal parameter values, regardless of the optimization strategies and momentum employed.

%By directly training the VAE with the PPGC model $\mathcal{M}$, in loop,  %process as described in (\ref{eq:direct_param_pred}), we observe that the generative model converges to a suboptimal point as the gradient stops flowing through the generator model. Although $M$ is continuous and differentiable, the sinusoidal and complex exponential terms in (\ref{eq:array_responses}) introduce periodicity in the PPGC model when observed as a function of the parameters $\theta_a,\theta_d$. As a result, the model quickly converges to one of the many local minimas spread across the entire landscape that have a large optimality gap, as seen by the loss curves of one of the training realizations in Fig. \ref{fig:nonconvex_surface}.

\subsection{Linearized Reformulation of the Physics Model}

\label{sec:pred_matrix}
To overcome the challenges posed by the PPGC model, while maintaining the key underlying model features for channel generation, we relax the overall cost function by reformulating the channel generation model $\mathcal{M}$ by first discretizing the range of values that $\theta_a^p,\theta_d^p$ can take and then expressing the channel generation process as a weighted average of the resulting antenna array responses vectors $\textbf{a}_r(\theta^p_a) \textbf{a}_t(\theta^p_d)^H$. Different from directly estimating the channel parameters, this method utilizes dictionary-based channel generation, where the loss function is now a linear function of the learnable dictionary weights. This is further explained next.
%We mitigate these problems by making the following changes. First, instead of using the PPGC model $\mathcal{M}$ directly to produce the channel, we now consider a weighted response of discretized antenna array responses over a predefined range of angles. 

Let the range of angles  $\theta_a^p,\theta_d^p$, given by $[\theta_{\min},\theta_{\max}]$, be equally divided into $R$ intervals of width $\Delta_{\theta} = (\theta_{\max}-\theta_{\min})/R$. We pre-compute the outer product between the array response vectors, $\textbf{a}_r(\theta^p_a) \textbf{a}_t(\theta^p_d)^H$, at the discretized angle values and store them in a dictionary $\textbf{D}$. The dictionary $\textbf{D}$ has a total of $R^{2}$ elements, and we refer to it as the array response dictionary across the remainder of this work. Each element of the dictionary, $\textbf{D}_{i,j}~ \forall~ i,j \in \{1,R\}$, represents a $N_r \times N_t$ matrix, and is given by

\begin{equation}
\label{eq:calc_array_dict}
    \textbf{D}_{i,j} = \textbf{a}_r(\theta_i) \textbf{a}_t(\theta_j)^H,
\end{equation} 
where $\big\{\theta_k := \theta_{\min}+k(\Delta_{\theta})|~  k \in \{i,j\}\big\}$.
Thus, each element of the array response dictionary is the combination of antenna array responses at the transmitter and receiver for certain values of the angle of arrival and angle of departure. The angles associated with neighboring elements of the dictionary $\textbf{D}_{i,j},\textbf{D}_{i+1,j}$ or $\textbf{D}_{i,j},\textbf{D}_{i,j+1}$ differ by a value of $\Delta_{\theta}$ for the angle of arrival or the angle of departure respectively.
The relaxed PPGC model can now be expressed as,
\begin{equation}
\label{eq:lincombi_channel}
    \textbf{H} = \sum_{i=1}^R\sum_{j=1}^R \textbf{W}_{i,j}\textbf{D}_{i,j},
\end{equation}
where, the channel generation is parametrized by the gain matrix $\textbf{W} \in \mathbb{R}^{R \times R}$, instead of the parameters $\textbf{s}$, and the channel is constructed by the element-wise product between $\textbf{W}$ and $\textbf{D}$. 
%The accuracy of the proposed relaxed model is strongly tied to the number of discretization intervals, $R$. As $R$ increases, the relaxed model in \eqref{eq:lincombi_channel} becomes closer to the model in \eqref{} 

By making these changes in the pipeline, we now model the output channel $\textbf{H}$ as a linear function of the gain matrix $\textbf{W}$, mitigating the issues arising from the non-convexity of the PPGC model $\mathcal{M}$ as seen in Sec.~\ref{sec:pred_params}. Thus, we now use the generative model to predict $\textbf{W}$ instead of $\textbf{s}$. Now, as the total number of paths $P$ is typically small, as mentioned in Sec.~\ref{sec:channel_model}, only a few of the entries in the $\textbf{W}$ are expected to be non-zero.

\textbf{Remark 1 :} It is to be noted that (\ref{eq:channel_sum}) and (\ref{eq:lincombi_channel}) both produce valid channel matrices. For a suitably high value of $R$, any channel $\textbf{H}$ can be approximated by (\ref{eq:lincombi_channel}) using a suitable sparse $\textbf{W}$ with $P$ non-zero values. For such a $\textbf{W}$, (\ref{eq:lincombi_channel}) is equivalent to (\ref{eq:channel_sum}), and each non-zero value $\textbf{W}_{i,j}$ will correspond to the gain $g_p$ of one of the $P$ paths. Thus, for this linearized representation of the problem, we can interpret the non-zero values of the gain matrix $\textbf{W}$ as the path gains, and the angles associated with the corresponding array dictionary vector $\textbf{D}_{i,j}$ as the angle of arrival and departure of those paths. 

\subsection{Generative Model to Predict the Gain Matrix}

In order to utilize the generative model to predict the gain matrix, we change the training pipeline described in Sec. \ref{sec:pred_params} in the following ways. 

\begin{algorithm}[tb]
\caption{Generation of PPGC using Linearized Model}
\label{algo:lin_model}

\textbf{\textit{\textbf{Training :}}}\\
\textbf{Given :} Dataset of valid channels $\mathcal{D}$, VAE parametrized by $\phi_e,\phi_d$, Chosen range of angles $[\theta_{\min},\theta_{\max}]$, Resolution $R$, Latent dimension $Z$.
\begin{algorithmic}[1]
    \STATE Calculate the array response dictionary $\textbf{D}$ using (\ref{eq:calc_array_dict}) for all $i \leq R, j \leq R$.
    \STATE Sample a channel matrix $\textbf{H}$ from the dataset $\mathcal{D}$.
    \STATE Obtain gain matrix $\textbf{W}$ from $\textbf{H}$ using (\ref{eq:vae_encode}) and (\ref{eq:vae_decode_lin}).
    \STATE Obtain the predicted channel matrix $\hat{\textbf{H}}$ from $\textbf{W}$ using (\ref{eq:vae_sum_weight}). 
    \STATE Calculate the loss using (\ref{eq:vae_loss}) and update $\phi_e,\phi_d$
\end{algorithmic}

\textbf{\textit{\textbf{Generation :}}}\\
\textbf{Given :} Sample vector $\tilde{\textbf{z}} \in \mathbb{R}^{Z}$ from Multivariate unit Gaussian distribution $\mathcal{N}(0,\textbf{I})$
\begin{algorithmic}[1]
    \STATE Obtain gain matrix $\tilde{\textbf{W}}$ from $\tilde{\textbf{z}}$ using (\ref{eq:gen_gain_matrix})
    \STATE Obtain generated channel matrix $\tilde{\textbf{H}}$ from $\tilde{\textbf{W}}$ using (\ref{eq:gen_channel_matrix})
    %\STATE Obtain the parameter coordinates $(i_p,j_p)_{p=1}^P$ of $\tilde{\textbf{H}}$ using $\tilde{\textbf{W}}$ on (\ref{eq:max_filter}),(\ref{eq:peak_detection}).
    %\STATE Obtain gain of the $p$-th path as $g_p = \tilde{\textbf{W}}_{i_p,j_p}$. Obtain the angles of arrival and departure of the $p$-th path as the angles used to generate $\textbf{D}_{i_p,j_p}$
\end{algorithmic}
\end{algorithm}

\begin{figure}
    \centering
    \includegraphics[width=0.98\columnwidth]{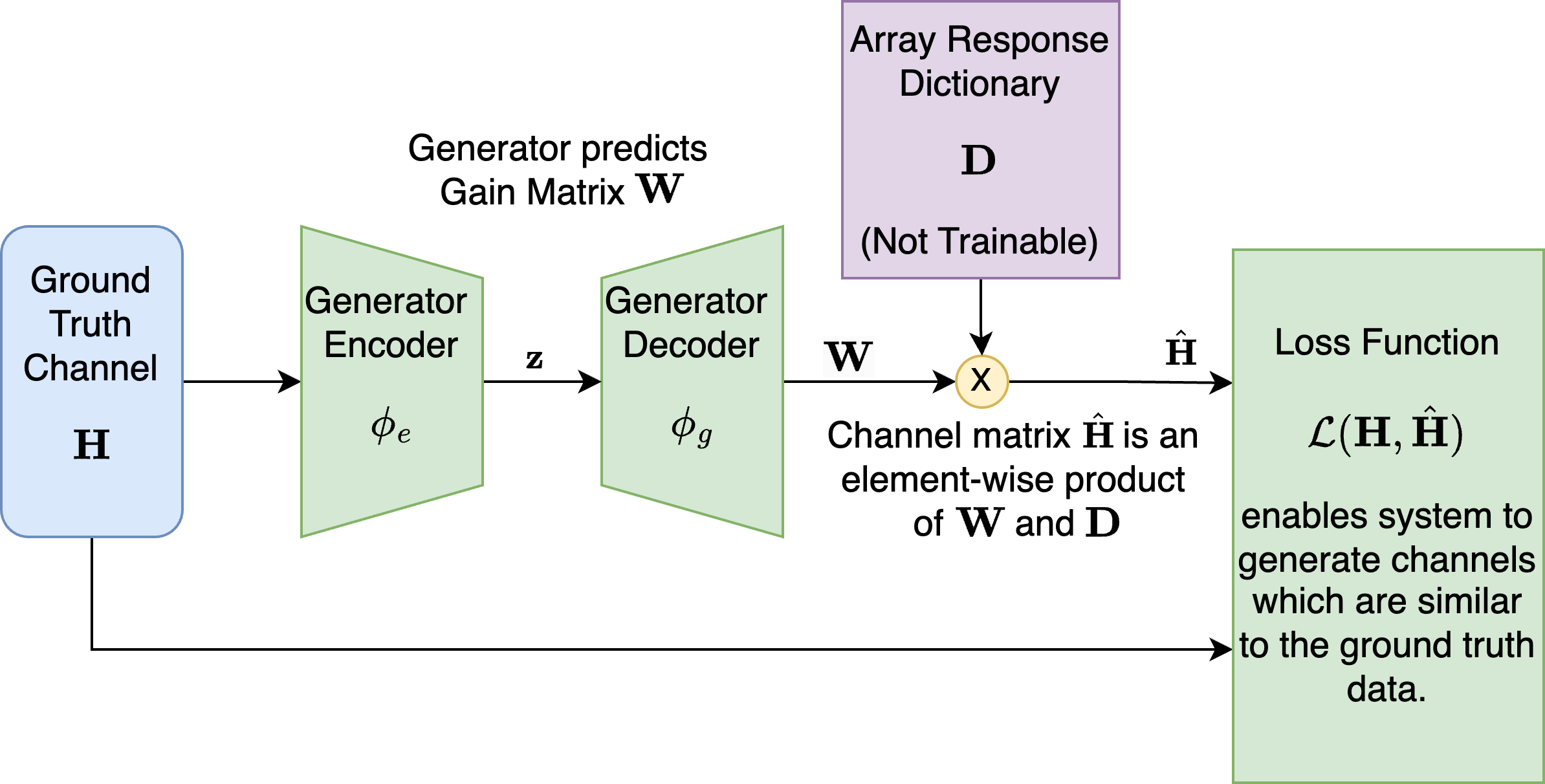}
    \caption{We relax the PPGC model by defining a discretized array response dictionary $\textbf{D}$ and using the generator to output the gain matrix $\textbf{W}$. The elementwise multiplication of $\textbf{W}$ and $\textbf{D}$ mimics the PPGC model process. This relaxation allows the flow of gradient through the generator, enabling it to converge to more suitable optima.}
    \label{fig:limmodel_train}
\end{figure}

The VAE decoder $g_{\phi_d}:\mathbb{R}^{Z} \rightarrow \mathbb{R}^{R \times R}$ now predicts the gain matrix $\textbf{W}$, where $R$ is the resolution of $\textbf{W}$. During training, the decoder $g_{\phi_d}$ takes in the latent vector $\textbf{z}$ as input and produces a gain matrix $\textbf{W}$ as
\begin{equation}
\label{eq:vae_decode_lin}
    \textbf{W} = g_{\phi_d}(\textbf{z}).
\end{equation}
The gain matrix is then used to generate the predicted channel by multiplying it with the array response dictionary as
\begin{equation}
\label{eq:vae_sum_weight}
    \hat{\textbf{H}} = \sum_{i=1}^{R}\sum_{i=1}^{R} \textbf{W}_{i,j} \textbf{D}_{i,j}.
\end{equation}
Now, we propose a modified system loss, that also accounts for the sparsity in the weight matrix $\mathbf{W}$. This is given by
\begin{equation}
\label{eq:vae_loss}
    \mathcal{L} = ||\textbf{H}-\hat{\textbf{H}}||_2^2 + \alpha_{D} \cdot \textsf{KL}(\textbf{z},\mathcal{N}(0,\textbf{I})) + \alpha_{S} \cdot ||\textbf{W}||_{1}.
\end{equation}
Here, the first two terms are identical to the ones used in (\ref{eq:vae_loss_direct}). The last term is the 1-norm of the gain matrix $\textbf{W}$, which encourages $\textbf{W}$ to be as sparse as possible, ensuring that $\textbf{W}$ does not select unrealistic combinations of multipaths that minimize the MSE. Thus, the overall reformulation can be interpreted as a sparse dictionary learning problem, with an additional KL divergence term.

\subsection{Inference and Sampling}

During the inference phase, we only use the generative decoder $g_{\phi_d}$. We sample a vector $\tilde{\textbf{z}} \sim \mathcal{N}(0,\textbf{I})$ and pass it through the decoder to produce a gain matrix $\tilde{\textbf{W}}$

\begin{equation}
\label{eq:gen_gain_matrix}
    \tilde{\textbf{W}} = g_{\phi_d}(\tilde{\textbf{z}})
\end{equation}
$\tilde{\textbf{W}}$ is then used to generate a synthetic channel, using the array response dictionary, as
\begin{equation}
\label{eq:gen_channel_matrix}
    \tilde{\textbf{H}} = \sum_{i=1}^{R}\sum_{i=1}^{R} \tilde{\textbf{W}}_{i,j} \textbf{D}_{i,j}.
\end{equation}
The generated channel $\tilde{\textbf{H}}$ is guaranteed to be a valid channel matrix as the mapping from the parameter space to the channel space is done based on a verified PPGC model. $\tilde{\textbf{H}}$ is also expected to be from the distribution of interest, as the second term in (\ref{eq:vae_loss}) ensures that the decoder of the generative model learns to map samples from the multivariate unit Gaussian distribution to a set of parameters that belong to the distribution of interest. 
A summary of the training and generative processes for the linearized representation is given in Algo. \ref{algo:lin_model}.

\section{Experimental Results}\label{sec:results}
\iffalse
\satya{\begin{itemize}
    \item Datasets used - Dataset generated using randomized parameters using ray-tracing simulator, DeepMIMO dataset.
    \item For synthetic data, gain, AoA and AoD ranges, number of antennas and number of samples.
    \item VAE architecture, optimizer, learning rate, batch size, number of epochs, latent dimension, weight matrix resolution.
    \item Compared against ChannelGAN.
    \item Exp : Our method predicts parameters accurately and within the ranges of the distribution (Synth).
    \item Exp : Our method predicts channels with high accuracy (DeepMIMO).
    \item Exp : Our method captures distinct, scenario-specific related parameter distributions and can be cross-evaluated on distinct datasets (DeepMIMO).
    \item Additional classification task downstream (?) Distinguish between BS 
\end{itemize}}
\fi

In this section, we analyze the performance of our method on wireless datasets generated based on user-defined parameter distributions as well as those that correspond to real-life scenarios, and compare our method against prior art baselines. We show the efficacy of our method in capturing the underlying parameter distributions, its ability to accurately generate synthetic channels, as well as its effectiveness of the proposed generative pipeline for downstream channel compression tasks.

For our PPGC model, we consider transmit and receive antennas $N_t = N_r = 16$. We consider six datasets. Five of the datasets are taken from the DeepMIMO framework \cite{deepmimo} that utilizes 3D ray tracing for generating channel datasets in different settings. Datasets corresponding to the following scenarios are used; (i) Two base stations in an outdoor intersection of two streets with blocking and reflecting surfaces. Here the channels corresponding to  base station 10 (BS10) and base station 11 (BS11) in the DeepMIMO framework have been considered; (ii) An indoor conference room, given by (Indoor); (iii) A section of downtown Boston, Massachusetts, USA, generated using the 5G model developed by RemCom \cite{remcom}, given by (Boston) and (iv) A section of the Arizona State University campus in Tempe, Arizona, USA, given by (ASU). The last is a user-defined dataset of size $20,000$ produced by sampling known distributions of parameters and using those parameters with the PPGC model $\mathcal{M}$ to produce a dataset (User Defined). More information regarding the datasets is provided in Appendix Sec. B. All datasets are split $80/20$ for training and testing respectively. 
%and all candidate UE positions in a grid of height $20$m and width $410$m starting from $(250,390)$m. The datasets for base stations $10,11$ consist of $20,004$ and $18,053$ datapoints respectively. 

\begingroup
\setlength{\tabcolsep}{3pt} % Default value: 6pt
\renewcommand{\arraystretch}{1.1} % Default value: 1
\begin{table*}[t]
    \small
    \centering
    \begin{tabular}{|c||c c c c|c c c c|c c c c| c c c c|}
    \hline
       \multirow{2}{2em}{Train} & \multicolumn{4}{|c|}{Testing $R_{10}$} & \multicolumn{4}{|c|}{Testing $R_{11}$} & \multicolumn{4}{|c|}{Testing $G_{10}$} & \multicolumn{4}{|c|}{Testing $G_{11}$} \\
        \cline{2-17}
    & Ours & CGAN & DUNet & CVAE & Ours & CGAN & DUNet & CVAE & Ours & CGAN & DUNet & CVAE & Ours & CGAN & DUNet & CVAE  \\
    \hline
    
    $R_{10}$ & 0.02 & 0.02 & 0.02 &0.02& \lightgray{1.35} & \lightgray{1.35} & \lightgray{1.35} & \lightgray{1.35} & \textbf{0.06} & 1.29 & 0.46 &0.55& \lightgray{1.36} & \lightgray{1.37} & \lightgray{1.18} & \lightgray{1.07}\\
    \hline
    
    $R_{11}$ & \lightgray{1.14} & \lightgray{1.14} & \lightgray{1.14} & \lightgray{1.14} & 0.06 & 0.06 & 0.06 &0.06& \lightgray{1.21} & \lightgray{1.51} & \lightgray{1.14} & \lightgray{1.04}& \textbf{0.25} & 0.45 &  0.42 & 0.72\\
    \hline
    
    $G_{10}$ & \textbf{0.05} & 0.77 & 0.19 &0.85& \lightgray{1.41} & \lightgray{0.97} & \lightgray{1.04} & \lightgray{1.49} & 0.01 & 0.09 & 0.03 &0.09& \lightgray{1.5} & \lightgray{0.92} & \lightgray{1.17} & \lightgray{1.1}\\
    \hline
    $G_{11}$ & \lightgray{1.1} & \lightgray{1.37} & \lightgray{1.02} & \lightgray{1.04} & \textbf{0.14} & 0.56 & 0.37 &0.52& \lightgray{1.31} & \lightgray{1.94} & \lightgray{1.33} & \lightgray{1.15} & 0.01 & 0.02 & 0.01 & 0.02\\
    \hline

    \end{tabular}
    \caption{NMSE loss for downstream compression tasks using different pairs of training and testing datasets. When compression models are trained on real data and evaluated on generated data (Rows $1$,$2$) and vice versa (Rows $3$,$4$), our method records lower NMSE for corresponding real-generated dataset pairs, indicating that the data generated by our method is more similar to the real channel data. }
    \label{tab:cross_eval}
\end{table*}
\endgroup

We use a VAE as the generative model, $(g_{\phi_e},g_{\phi_d})$, which is trained using the Adam optimizer with a learning rate of $1e^{-3}$. For more details regarding the model architecture, please refer to Appendix Sec. A. We compare our model against  ChannelGAN (CGAN)\cite{channelgan}, the DUNet diffusion model (DUNet) \cite{diffusion_models_for_channels} and a VAE version of CSINet \cite{csinet}. Models are trained for $300$ epochs with a batch size of $256$. For our model, we use resolution $R = 64$ and latent dimension $z = 64$.

\begin{figure}
    \centering
    \includegraphics[width=0.49\columnwidth]{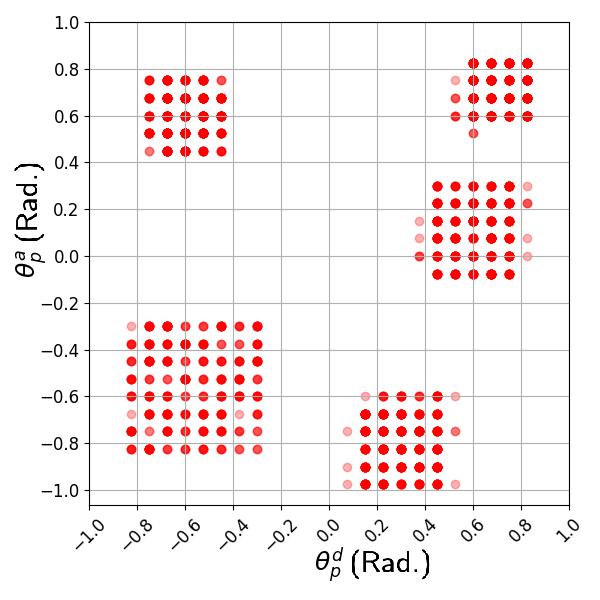}
    \centering
    \includegraphics[width=0.49\columnwidth]{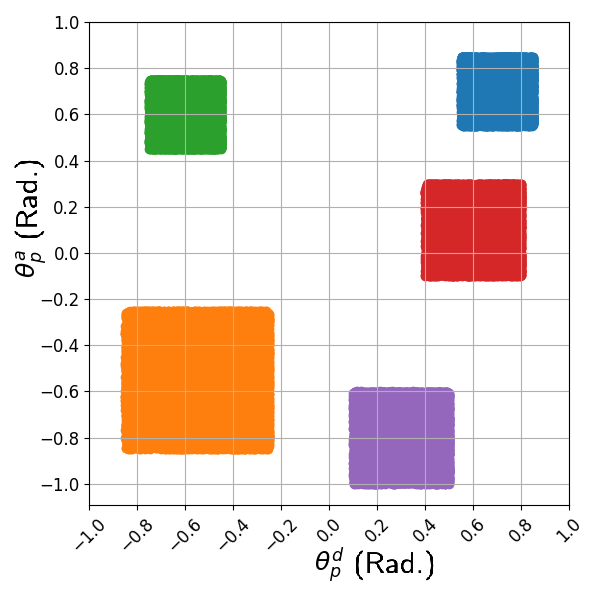}
    \label{fig:enter-label}
    \caption{The distributions of angles of arrival and departure $(\theta_a^p,\theta_d^p)$ captured by our method (Left) match the underlying distributions of the training dataset (Right). Each color corresponds to a distinct path.}
    \label{fig:param_comp}
\end{figure}
\subsection{Prediction of Parameters} 
In this experiment, we evaluate the accuracy of our system in capturing the underlying distribution of parameters associated with a given dataset. We first define a distribution across the parameters $[g_p,\theta_p^a,\theta_p^d]_{p=1}^P$ and generate a dataset $\textbf{D}$ of channel matrices by sampling parameters from this distribution and passing them through the model $\mathcal{M}$. We train our model on the dataset $\textbf{D}$ and compare the distribution of predicted parameters $[\tilde{g}_p,\tilde{\theta}_p^a,\tilde{\theta}_p^d]_{p=1}^P$ with that of the true parameters $[g_p,\theta_p^a,\theta_p^d]_{p=1}^P$. 

In Fig. \ref{fig:param_comp}, we observe that our method can accurately capture the distributions of the angles of arrival and departure for each path. The discretized array response dictionary results in grids of generated angles of arrival and departure, which align with the distribution of parameters used to generate the channels. This shows that our model training and parameter extraction methods can be used to determine the distributions of parameters of input channels without requiring labeled data.

\begin{figure*}[t!]
    \centering
    \begin{subfigure}{0.32\textwidth}
        \centering
        \includegraphics[height=1.7in]{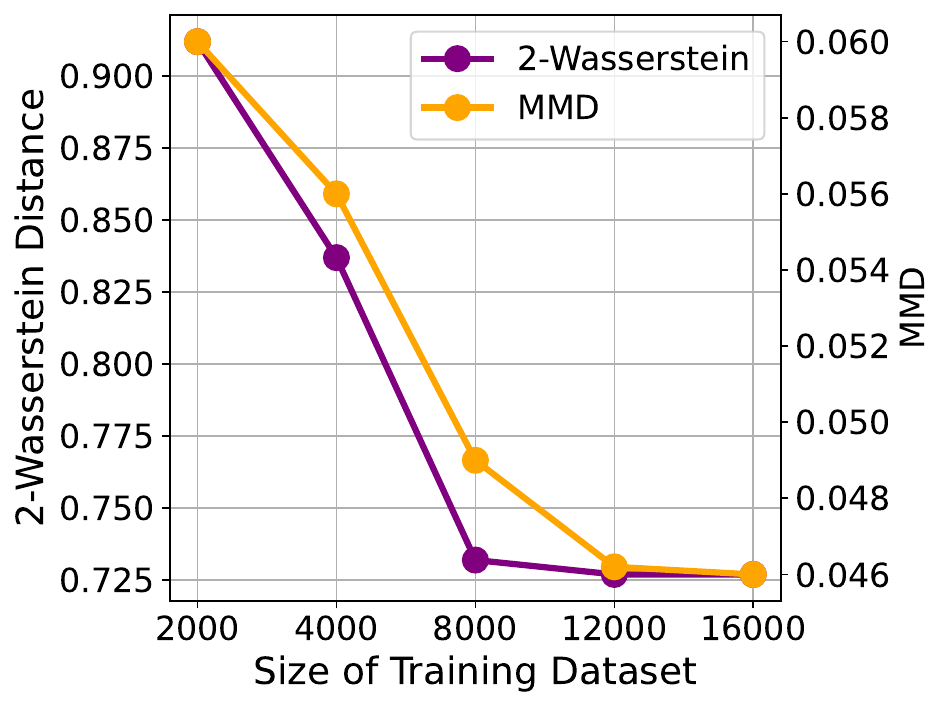}
        \caption{}
        \label{fig:abl_num_dp}
    \end{subfigure}
    \begin{subfigure}{0.32\textwidth}
        \centering
        \includegraphics[height=1.7in]{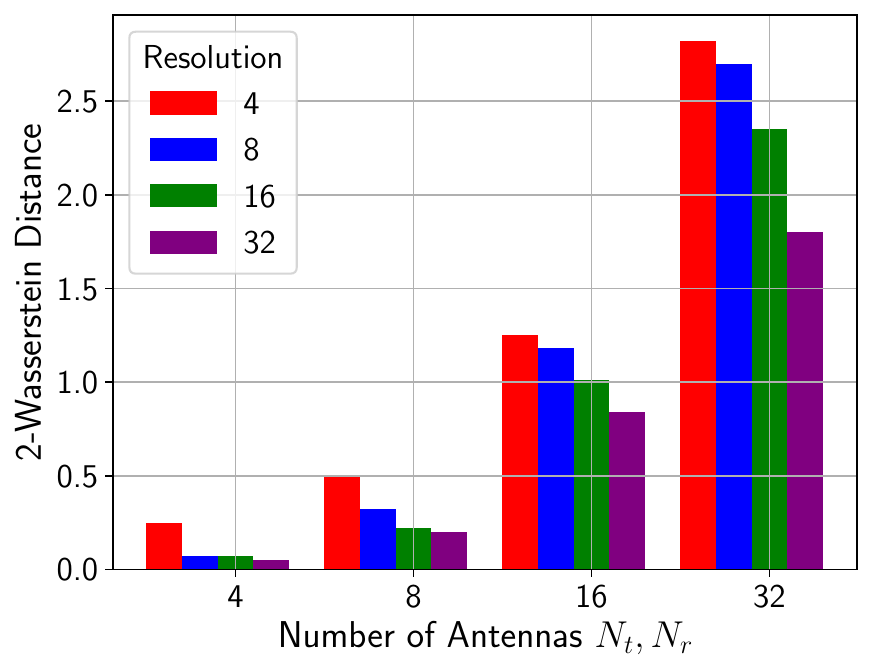}
        \caption{}
        \label{fig:abl_resolution}
    \end{subfigure}
    \begin{subfigure}{0.32\textwidth}
        \centering
        \includegraphics[height=1.7in]{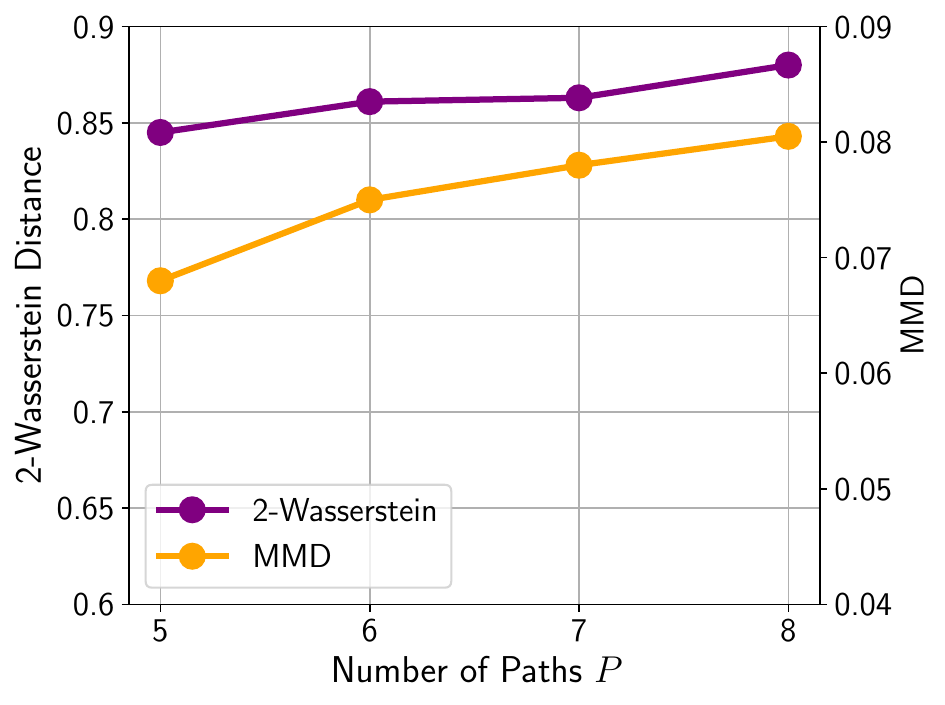}
        \caption{}
        \label{fig:abl_num_paths}
    \end{subfigure}
    \caption{(a) Our method can generate samples with a high degree of fidelity in terms of 2-Wasserstein distance and MMD, even with a training dataset of around $50\%$ of the size of the training dataset. (b) For a physics model $\mathcal{M}$, as the number of antennas $(N_r,N_t)$ increase, an increase in the resolution ($R$) of the gain matrix $\textbf{W}$ results in a higher degree of fidelity with the input data distribution in terms of the 2-Wasserstein distance, as the highest possible precision with which parameters can be estimated is dependent on the number of antennas. (c) The performance of our method is consistent across a varying number of paths $P$, as the generative process is independent of $P$, relying on the loss function to balance reconstruction fidelity and the identified number of paths, given by the number of non-zero values in the gain matrix \textbf{W}.}
\end{figure*}

\subsection{Reconstruction of Channels}
In Table \ref{tab:channel_dist}, we analyze the ability of our method to capture the distribution of channel matrices compared to baseline methods. We generate $3000$ synthetic channel matrices and compare the 2-Wasserstein distance \cite{2wasserstein} and Maximum Mean Discrepancy \cite{mmd} between the distribution of the generated channels and the true channels.

%We observe that the channels generated by our method in Fig. \satya{FIG}.
The channels generated by our method are closer to the distribution of true channels than those generated by the ChannelGAN baseline by up to $4 \times$. This shows that our method can generate more realistic channel data as compared to baselines.

\begingroup
\setlength{\tabcolsep}{1.2pt} % Default value: 6pt
\renewcommand{\arraystretch}{1.1} % Default value: 1
\begin{table}[t]
    \small
    \centering
    \begin{tabular}{|c||c|c|c|c|c|c|c|c|}
    \hline
       & \multicolumn{4}{|c|}{2-Wasserstein Distance}  & \multicolumn{4}{|c|}{MMD}\\
      \hline
      Dataset   & Ours & CGAN & DUNet & CVAE & Ours & CGAN & DUNet & CVAE\\
      \hline
      Boston  & \textbf{0.475} & 0.84 & 0.687 & 0.87 & \textbf{0.013} & 0.045 & 0.052 & 0.095 \\
       ASU  & \textbf{0.283} & 0.932 & 0.641 & 1.095 & \textbf{0.012} & 0.169 & 0.02 & 0.592 \\
       Indoor & \textbf{0.17} & 0.339 & 0.529 & 1.802 & \textbf{0.005} & 0.003 & 0.02 & 0.375 \\
       BS10  & \textbf{0.656} & 1.632 & 0.857 & 1.837 & \textbf{0.072} & 0.317 & 0.108& 0.414\\
       BS11  & \textbf{0.267} & 0.882 & 0.463 & 1.12 &\textbf{0.016} & 0.086 & 0.033& 0.132\\
       \hline
    \end{tabular}
    \caption{The distribution of channels modelled by our method is more similar to the real distribution compared to baselines in terms of 2-Wasserstein distance and Maximum Mean Discrepancy (MMD).}
    \label{tab:channel_dist}
    \vspace{-3mm}
\end{table}
\endgroup

\subsection{Cross Evaluation Between Distinct Datasets}

In this experiment, we analyze the ability of our method to learn distinct channel distributions based on the cross-evaluation of models trained on different scenarios in the context of a downstream channel compression task. In practical wireless deployments, effective channel compression is crucial in mitigating feedback overheads that arise in systems that frequently exchange channel data to optimize communication efficiency \cite{csinet,csicomp}.

We consider two channel datasets $R_{10},R_{11}$ from the DeepMIMO Outdoor scenario, generated from base stations $10$ and $11$ respectively. We train an independent instance of the generative model on each dataset and generate synthetic datasets of size $20,000$, given by $G_{10},G_{11}$ respectively. We then train independent instances of the CSINet channel compression model \cite{csinet} on each set. We perform cross evaluation considering all pairwise combinations of training and testing datasets and calculate the test NMSE given in Table \ref{tab:cross_eval}.

Now, a model trained on $G_{10}$ should generalize well to $R_{10}$, and vice versa for a model trained on $G_{11}$. In Table \ref{tab:cross_eval}, we observe that a compression model trained on data generated by our model follows the aforementioned rules, indicating that our method can capture the distinctions between two different datasets and generate distinct channel data samples.  

%\begin{figure}
%    \centering
%    \includegraphics[width=0.95\columnwidth]{images/abl_num_dp.png}
%    \caption{Our method can generate samples with a high degree of fidelity with a smaller subset of training data. Even with a training dataset of around $50\%$ of the size of the training dataset, our method provides comparable results in terms of 2-Wasserstein distance and MMD.}
%    \label{fig:abl_num_dp}
%\end{figure}

%\begin{figure}
%    \centering
%    \includegraphics[width=0.95\columnwidth]{images/abl_resolution.png}
%    \caption{}
%    \label{fig:abl_num_dp}
%\end{figure}

\subsection{Effect of Varying Size of Training Dataset}
In this experiment, we train our model on datasets of varying sizes, choosing a subset of the original dataset of the appropriate size. We use the DeepMIMO dataset BS10 for this experiment, which consists of $16,000$ datapoints.

In Fig. \ref{fig:abl_num_dp}, we observe that our method is able to provide a similar level of performance even when the size of the training data is reduced by up to $\sim 40\%$. This is because our model predicts distributions in the parameter space, which are less complex than the distributions in the channel space, even a small number of datapoints can capture the distribution related characteristics of the input channels. In a practical deployment, this translates to significant savings in terms of the resources deployed to acquire channel data.

\subsection{Effect of Varying Resolution}
In this experiment, we observe the joint effect of changing the resolution $R$ of the gain matrix $\textbf{W}$, and the number of antennas $N_r,N_t$ used by the physics model $\mathcal{M}$. For this experiment, we use the DeepMIMO dataset for base station $10$. 

In Fig. \ref{fig:abl_resolution}, we observe that for a model with fewer antennas $(N_r=N_t=4/8)$, an increase in resolution $R$ offers diminishing improvements. This is because the highest level of granularity at which parameters can be estimated is dependent on the number of antennas, and if the resolution is increased beyond this point, there is little to no change in the performance. However, for a model with more antennas $(N_r=N_t=16/32)$, the added dimensionality of the channel matrices allows the model to identify angles of arrival and departure, $(\theta_a^p,\theta_d^p)_{p=1}^P$ with greater precision. Thus, in such cases, increasing the resolution of the weight matrix $\textbf{W}$ improves the performance of the model considerably.

\subsection{Effect of Increasing Number of Paths}
In this experiment, we observe the effect of increasing the number of paths $P$ in a channel. We consider the user defined dataset and add additional paths where needed as follows. Path $6$ is sampled from $\theta_p^a \sim \mathcal{U}(0.4,0.8)/ \theta_p^a \sim \mathcal{U}(0.1,0.3)$, Path 7 is sampled from $\theta_p^a \sim \mathcal{U}(0.6,1.0)/ \theta_p^a \sim \mathcal{U}(-0.3,-0.1)$, and Path $8$ is sampled from $\theta_p^a \sim \mathcal{U}(-0.3,0.9)/ \theta_p^a \sim \mathcal{U}(0.6,1.0)$. $g_p \sim \mathcal{U}(0.001,0.01)$ for all paths.

In Fig. \ref{fig:abl_num_paths}, we observe that the performance of our generative pipeline remains consistent across a varying number of paths $P$. This is because our model is independent of $P$, and leverages the formulation of the loss function in (\ref{eq:vae_loss}) to balance the reconstruction accuracy and the number of non-zero output values, which dictates the number of identified paths. This is encapsulated by the last term in (\ref{eq:vae_loss}), which enforces output sparsity. The hyperparameter $\alpha_S$ in the loss function can thereby be tuned by observing the reconstruction loss. Thus, our method can adapt to a range of values for the number of paths by finding a suitable balance between the NMSE and the L1 regularization loss such that the number of non-zero values are proportional to the number of paths.

%larger resolution also results in a larger model, as seen from the last layer of the decoder in Fig. \ref{fig:model_arch}. As a result, the additional compute resources required for an architecture with a higher resolution are unnecessarily used.

\section{Conclusion}\label{sec:conclusion}
In this paper, we developed a generative pipeline that leverages a PPGC model for parametrized channel generation. We tackle the extreme non-linearity in the PPGC model by developing a dictionary-based relaxation of the PPGC model and learning a sparse gain matrix whose non-zero values denote the parameters of the associated paths.
We empirically show that our method effectively captures path-specific parameter distributions for a given dataset of channel matrices and outperforms prior art in terms of 2-Wasserstein distance and MMD.
Our work can be extended to 3-dimensional scenarios with angles of elevation and additional parameters such as path delay. 

\section{Acknowledgements}

This work was completed during Satyavrat Wagle's internship at InterDigital Communications.

Prof. Christopher G. Brinton and David J. Love helped to draft this article under support from the Office of Naval Research (ONR) grant N00014-21-1-2472, and the National Science Foundation (NSF) grants CNS-2212565 and ITE-2326898. Prof. Christopher G. Brinton was also supported in part by the Defense Advanced Research Projects Agency (DARPA) grant D22AP00168.

\section{Code Release}

The code to run all experiments in this paper can be found at \textsf{https://github.com/akshaymalhotraidcc/Physics-Informed-Generative-Wireless-Channel-Modeling}.

\bibliographystyle{named}
\bibliography{refs}
\end{document}